\documentclass[aps, prl, twocolumn,superscriptaddress,showpacs,preprintnumbers,amsmath,amssymb]{revtex4}

\usepackage{graphicx}
\usepackage{dcolumn}
\usepackage{bm}

\newcommand{\ket}[1]{\ensuremath{\left|{#1}\right \rangle}}

\begin{document}


\title{All-Optical Ultrafast Control and Read-Out of a Single Negatively Charged Self-Assembled InAs Quantum Dot}
\author{Erik D. Kim}
\author{Katherine Smirl}
\author{Yanwen Wu}
\author{A. Amo}
\author{Xiaodong Xu}
\author{D. G. Steel}
 \email{dst@umich.edu}
\affiliation{The H. M. Randall Laboratory of Physics, The University of Michigan, Ann Arbor, MI 48109, USA}
\author{A. S. Bracker}
\author{D. Gammon}
\affiliation{The Naval Research Laboratory, Washington D.C. 20375, USA}
\author{L. J. Sham}
\affiliation{Department of Physics, The University of California, San Diego, La Jolla, CA, 92093-0319, USA}

\date{\today}

\begin{abstract}
We demonstrate the all-optical ultrafast manipulation and read-out of optical transitions in a single negatively charged self-assembled InAs quantum dot, an important step towards ultrafast control of the resident spin. Experiments performed at zero magnetic field show the excitation and decay of the trion (negatively charged exciton) as well as Rabi oscillations between the electron and trion states. Application of a DC magnetic field perpendicular to the growth axis of the dot enables observation of a complex quantum beat structure produced by independent precession of the ground state electron and the excited state heavy hole spins.

\end{abstract}

\pacs{78.67.Hc, 71.35.Pq, 42.50.Md, 42.50.Hz}

\maketitle

Quantum dots (QDs) containing a single spin have been the focus of intensive efforts over the last several years to demonstrate their viability for use in quantum computing schemes \cite{LossPRA, ImamogluPRL, HansonRMP, Saikinarx}. Major advances in this effort include the observation of both a long spin lifetime \cite{ElzermanNAT, KroutvarNAT, HansonPRL} and a long spin coherence time \cite{JRPettaSCI, BayerSCI, AwschNatPhys}. Crucial to all quantum computing schemes proposing the use of QD-confined spins is the ability not only to carry out arbitrary coherent manipulations of the spin qubit, but to do so on a time-scale much shorter than the spin decoherence time \cite{DiViFortsch}. These capabilities are possible with ultrafast optical excitation of a three level system containing the spin states of an electron confined in a QD.

Considerable progress towards optically manipulating QD-confined spins on ultrafast timescales has been made in interface fluctuation dots, where ensemble studies have demonstrated the generation and read-out of electron spin coherence \cite{GuruPRL} as well as partial rotations of the electron spin \cite{WuPRL}. At the single dot level, time-resolved Kerr-rotation measurements have recently shown the optical read-out \cite{AwschNatPhys} and partial rotation \cite{AwschSci} of an electron spin in an interface fluctuation dot integrated in an optical cavity. Despite these successes, the weak lateral confinement in these dots \cite{GammonPRL} and the inability to grow them in patterned arrays pose substantial challenges for the implementation of interface fluctuation dots in a practical quantum computing architecture.

Self-assembled QDs provide an attractive alternative to interface fluctuation dots due to their stronger spatial confinement \cite{PetroffPRB} and their ability to be organized in 2D and 3D lattices during growth \cite{LeeAPL}. Transient optical studies of these dots, however, are made difficult by their optical dipole moments, which are 1-2 orders of magnitude smaller than those of interface fluctuation dots \cite{GuestPRB}. This difficulty necessitates extremely low noise levels for pulsed optical measurements. As a result, most transient optical studies of charged self-assembled QDs have measured the optical response of an ensemble of these dots to obtain larger signals \cite{BayerSCI}, or, in the case of single dot studies, the photoluminescence \cite{YamNat} or the photocurrent of a dot embedded in a photodiode structure \cite{ZrennerNat}. Though the last approach has demonstrated ultrafast preparation and read-out of a single QD-confined hole spin \cite{AJPRL}, the fast tunneling rates of electrons and holes required for photocurrent read-out places a limit on the operation time of potential spin qubits in these systems. Hence, optical read-out through time integrated pump-probe spectroscopy (as presented in this work) or through optical cycling \cite{DannyPRL} provides an alternative approach that may be more useful for some applications.

In this Report, we demonstrate the coherent transient optical control and read-out of the states of a single self-assembled QD via resonant ultrafast optical pulses. Our read-out technique effectively measures the difference between the occupation probabilities of the optically coupled levels at a particular time, allowing for the observation of transient phenomena. At zero magnetic field, pump-probe studies show the optical excitation and decay of the trion (negatively charged exciton) and are used to demonstrate coherent control of the electron-trion transition. Applying an external DC magnetic field perpendicular to the growth axis (in the Voigt geometry) mixes the electron spin states and the hole spin states of the trion, lifting the Kramer's degeneracy for each spin. This allows for two-photon Raman type excitations that create quantum coherence between the corresponding spin states of the electron and the trion. As a result of the coherence, we are able to observe the precession of electron and heavy-hole spins in a single QD.

The sample under study contains a single layer of self-assembled InAs QDs grown on a GaAs substrate in a Schottky diode structure that enables selective control of the number of electrons in a given dot via an externally applied bias voltage \cite{WarePRL}. An Al mask containing apertures approximately 1 $\mu$m in diameter allows for transition energy selective optical excitation of single QDs. The sample is placed in a magnetic He-flow cryostat enabling operating temperatures of approximately 5 K. A tunable 76 MHz Ti:Sapph laser generating mode-locked sech pulses approximately 2 ps in width is used as the excitation source. QDs are chosen such that their optical transition energies are separated from those of other dots by an amount greater than the pulse optical bandwidth.

\begin{figure}[b]
\includegraphics{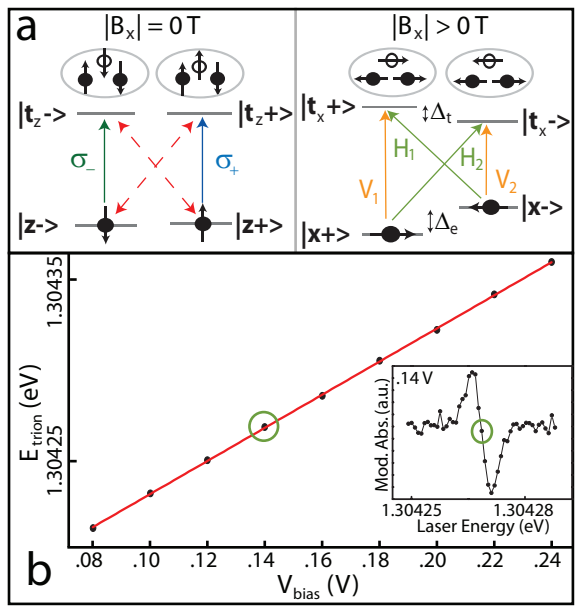}
\caption{\label{fig1} (color online). (a) Energy level diagrams for a single InAs dot charged with an electron for both zero and nonzero external magnetic fields. The dashed red arrows indicate optically forbidden transitions. (b) $\text{E}_{\text{trion}}$ as a function of $\text{V}_{\text{bias}}$ at zero magnetic field. The red curve is a linear fit of the data, indicating a linear Stark shift. The inset shows a modulation absorption scan taken at a sample bias of $.14$ V for small modulation amplitude, giving the derivative of the trion absorption line \cite{AlenAPL}. The energy of the zero crossing point (green circle) corresponds to the trion transition energy.}
\end{figure}

The lowest-lying energy levels for a single InAs dot containing an electron are shown in Fig.~\ref{fig1}(a), with and without an externally applied magnetic field \cite{XuPRL}. At zero magnetic field, the levels are quantized along $\hat{z}$ and form two degenerate, circularly cross polarized two-level systems. For the case of a magnetic field applied along $\hat{x}$, the QD levels form a system where the electron spin states and the trion spin states are separated by their respective Zeeman energies and state mixing allows each electron spin state to be optically coupled to both trion states by linearly polarized selection rules. In both cases, the QD levels may be tuned via the DC Stark effect by varying the sample bias voltage, $\text{V}_{\text{bias}}$. As $\text{V}_{\text{bias}}$ also controls the number of electrons within the dot, there exists a limited voltage range in which the trion transition energy, $\text{E}_{\text{trion}}$, may be shifted without altering the charge state of the dot. Fig.~\ref{fig1}(b) shows the range of transition energies covered by the trion voltage range as determined by Stark-shift modulation absorption studies \cite{AlenAPL}.

An optical pulse incident along the sample growth axis and resonant with one of the trion transitions in a single dot provides a means of probing the dot by generating an optical polarization within it. This polarization radiates an optical field proportional to the difference in occupation probabilities encountered by the pulse between the optically coupled QD levels. Since the radiated field reflects the state of the dot at the arrival of the pulse, measurements of its amplitude enable time-resolved single dot studies. In what follows, we consider pump-probe studies where the probe pulse is used to effectively monitor the time-evolution of the QD after excitation by the pump pulse.

To isolate the probe-generated field amplitude, we perform time-averaged measurements of the interference between the radiated field and the transmitted probe on a square-law detector. We also implement a bias voltage scheme that modulates the probe-induced optical polarization, allowing for the use of phase-sensitive detection with a lock-in amplifier to improve measurement signal-to-noise. Pump-probe delays are kept under 1.2 ns so that the optically generated trion population and coherence decays before the subsequent pulse pair, given respective decay times of approximately 1 ns and 2 ns \cite{WoggonPRB, XuPRL}. In studies with an externally applied magnetic field, it may be possible to generate electron spin coherence that persists beyond the repetition period of the laser. This will give rise to a signal for negative values of the pump-probe delay where the probe measures the effect of the pump from the previous pulse pair, as observed in the studies of Ref.~\cite{BayerSCI}.

Studies performed without an externally applied magnetic field enable observation of trion excitation and decay and are used to demonstrate coherent control of the electron-trion transition. Linearly cross polarized pump and probe pulse trains are used to enable post-sample filtering of the pump. In thermal equilibrium at 5 K, the population in a singly-charged QD is equally distributed between the two spin ground states. For a sample bias within the voltage range of the trion, the incident $H$-polarized pump pulse interacts equally with both transitions in the dot assuming equal transition dipole moments. Any trion population generated by the pump then decays exponentially to the corresponding ground state at the trion relaxation rate, $\Gamma_t$. Due to the polarization of the subsequent probe pulse, the occupation probability difference encountered by the probe in each transition contributes equally to the lock-in detected signal. We thus consider the time-dependent occupation probability difference in just one of the transitions.

\begin{figure}
\includegraphics{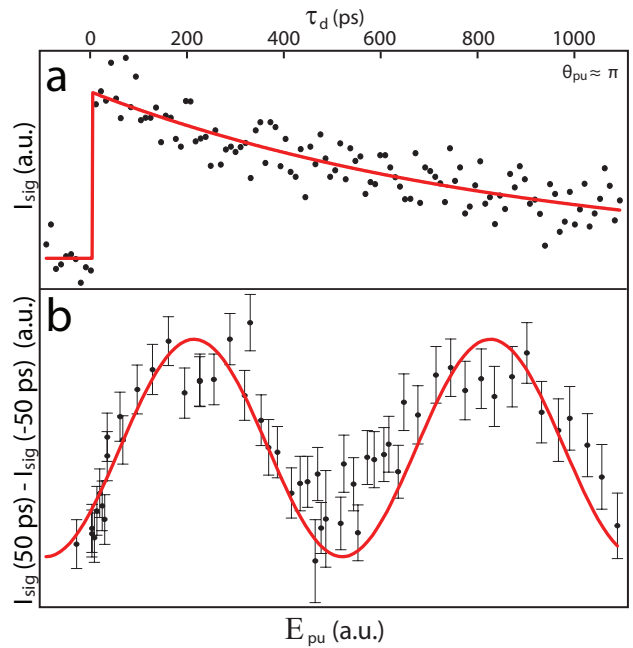}
\caption{\label{fig2} (color online). (a) Zero-field $\text{I}_{\text{sig}}$ measurements showing the generation of trion population at zero delay followed by an exponential decay. The solid line is a fit of the data using the second term of Eq.~(\ref{eq1}). (b) Plot of the difference between $\text{I}_{\text{sig}}$ values at 50 ps and at -50 ps as a function of pump electric field amplitude, keeping the probe pulse area at approximately $\frac{2\pi}{5}$. The data show two Rabi oscillations where the solid line is a fit to an oscillation of the form of Eq.~(\ref{eq1}).}
\end{figure}

An analytical expression for the occupation probability difference may be obtained by solving for the matrix elements of the density operator, $\hat{\rho}$, for the two-level system formed by one of the trion transitions. Taking the trion dephasing time to be much longer than the pump pulsewidth, the system is first solved under excitation by the incident pump of pulse area $\theta_{pu}$, defined as $\theta_{pu} = \frac{1}{\hbar}\int_{-\infty}^{\infty}dt'\boldsymbol\mu\cdot\mathbf{E}_{pu}(t')$ where $\boldsymbol\mu$ is the transition dipole moment and $\mathbf{E}_{pu}(t)$ is the electric field of the pump pulse. Solution values after excitation are then used as the initial conditions for the freely decaying system. The difference between the diagonal elements of the resultant density matrix, $\Delta \rho$, corresponds to the difference in trion and electron occupation probabilities and has the form
\begin{align}
\Delta\rho(t) = - 1/2 + \Theta(t)\sin^2(\theta_{pu}/2)e^{-\Gamma_t t} \label{eq1}
\end{align}
where $-1/2$ is the occupation probability difference prior to excitation in each transition, $\Theta$ is the Heaviside function and $\sin^2(\theta_{pu}/2)$ is twice the amount of optically generated trion population immediately after the pump. It is understood that Eq.~(\ref{eq1}) applies for absolute values of time greater than the pump pulse width. $\Delta\rho$ leads to the lock-in detected signal $\text{I}_{\text{sig}}$ and is plotted in Fig.~\ref{fig2}(a) as a function of the pump-probe delay, $\tau_d$. The predicted exponential decay of the signal is clearly evident. Fitting the data to the second term of Eq.~(\ref{eq1}) yields a trion lifetime of $855\pm74$ ps, consistent with values reported in separate time-resolved photoluminescence \cite{WoggonPRB} and trion linewidth \cite{XuPRL} measurements.

Coherent control of the electron-trion transition, a crucial requirement for two-photon Raman control of the electron spin, may be demonstrated by performing power-dependent measurements of $\text{I}_{\text{sig}}$. As the second term in Eq.~(\ref{eq1}) depends on the trion population generated by the pump, the amplitude of the exponential decay in delay scan measurements of $\text{I}_{\text{sig}}$ [Fig.~\ref{fig2}(a)] oscillates as a function of $\theta_{pu}$ due to pump-driven Rabi oscillations between the electron and trion states. To observe these Rabi oscillations, we take the difference between a positive and a negative delay measurement of $\text{I}_{\text{sig}}$ as a function of pump pulse area. These difference measurements allow us to observe Rabi oscillations while accounting for a power dependent offset in $\text{I}_{\text{sig}}$ that arises from pump leakage through the post-sample filtering setup. From Eq.~(\ref{eq1}), these difference measurements yield a signal proportional to $\Delta\rho(\tau_+) - \Delta\rho(\tau_-) = e^{-\Gamma_t\tau_+}\sin^2(\theta_{pu}/2)$, where $\tau_{+(-)}$ is a positive (negative) value of the pump-probe delay. Two Rabi oscillations are plotted in Fig.~\ref{fig2}(b) as a function of pump pulse amplitude for $\tau_{\pm} = \pm 50$ ps. An oscillatory fit of the data is given by the red curve in Fig.~\ref{fig2}(b), showing qualitative agreement between theory and experiment. From the oscillation frequency we extract a transition dipole moment of approximately 8 Debye. There is also an additional weak oscillation in the data that shows up noticeably at the peaks of the primary oscillations. This structure is likely the result of coupling to a nearby dot whose trion transition energy is within a few pulse widths \footnote{Inclusion of the nearby trion with parameters well within the range of those observed in these samples leads to an additional oscillation in the theoretically calculated Rabi oscillations, consistent with experimental results.}.

Application of an external DC magnetic field in the Voigt geometry lifts the Kramer's degeneracy for the electron states and the lowest-lying trion states, enabling observation of both electron and the heavy hole spin precession with the use of circularly cross polarized pump and probe pulse trains. The resonant $\sigma_{+}$ polarized pump pulse serves to generate Raman coherence between the spin states of both the electron and the heavy-hole via two-photon processes, thereby initializing oppositely oriented electron and heavy-hole spin polarization vectors along the optical axis. Due to the presence of the external magnetic field, the spin polarization vectors precess about $\hat{x}$ at rates determined by the electron and heavy-hole in-plane g-factors, which generally differ. Over the course of the precession, the magnitudes of the optically induced spin polarization vectors decay in time. The rate of decay for the electron is determined by electron spin dephasing time, while the rate of decay for the heavy-hole is determined primarily by the trion relaxation time. These decaying spin precessions manifest as damped oscillations in the occupation probability difference encountered by the time-delayed $\sigma_{-}$ polarized probe pulse in the $\ket{\text{z}-}$ to $\ket{\text{t}_\text{z}-}$ transition.

\begin{figure}[t]
\includegraphics{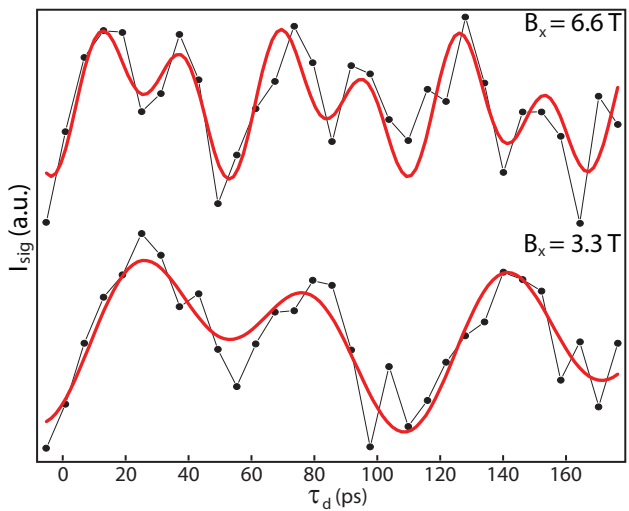} 
\caption{\label{fig3} (color online). $\text{I}_{\text{sig}}$ as a function of pump-probe delay for external magnetic field values of 3.3 T (bottom) and 6.6 T (top). The complicated quantum beat signatures in both cases indicate precessions of the electron and the heavy hole spins at their respective precession frequencies. As the pump-probe delay range is short compared to previously observed electron and heavy-hole coherence times, the data are fit to Eq.~(\ref{eq2}) taken in the limit of very long spin coherence times (red curves).}
\end{figure}

As with zero-field studies, an expression for the difference in occupation probabilities encountered by the probe may be obtained by solving for the density matrix elements, though here all four levels must be considered. Once again, we take the electron spin states to be completely mixed prior to excitation. In this case, $\Delta\rho$ is given by the difference between the diagonal matrix elements corresponding to the $\ket{\text{t}_\text{z}-}$ and $\ket{\text{z}-}$ states and has the form
\begin{equation}
\begin{split}
\Delta &\rho(t) = -1/2 + (1/4)\Theta(t)\sin^2\left(\theta_{pu}/2\right)e^{-2\Gamma_t t}\times \\
{} & \left[2 - e^{-t/T^{h*}_2}\cos(\omega_{h}t)- e^{2\Gamma_t t}e^{-t/T^{e*}_2}\cos(\omega_{e}t)\right] \label{eq2}
\end{split}
\end{equation}
where $-1/2$ is the occupation probability difference prior to excitation, $\omega_{e(h)}$ is the electron (heavy-hole) spin precession frequency and $T_{2}^{e(h)*}$ is the effective coherence time of the electron (heavy-hole) spin due to the fluctuating nuclear spin environment in the dot. $\text{I}_{\text{sig}}$ measurements at external magnetic field values of 3.3 T and 6.6 T are shown in Fig.~\ref{fig3}, exhibiting two-frequency modulations caused by the electron and heavy-hole spin precessions. The limited range of pump-probe delays in Fig.~\ref{fig3} prevents reliable extraction of the electron and heavy-hole spin coherence times, as this range is much shorter than previously measured electron spin coherence times in similar dots \cite{CPTXD} and anticipated heavy-hole spin coherence times in these systems \cite{HoleCohNatMat}. Hence, the data are simply fit to Eq.~(\ref{eq2}) assuming no spin dephasing (solid red curves of Fig.~\ref{fig3}). The agreement between the fitting and the data suggests spin coherence times that are much longer than the pump-probe delay values considered. From the fitting, we obtain electron and heavy-hole g-factor magnitudes $|g_e|$ and $|g_h|$ of $.378\pm.007$ and $.202\pm.006$, respectively, in agreement with separate frequency domain measurements we made of the electron and heavy-hole Zeeman splittings (data not shown).

We have shown the coherent manipulation and read-out of the optical transitions in a single self-assembled InAs dot using resonant picosecond optical pulses. Our approach provides a useful configuration for demonstrating the operations required to implement QD confined spins for quantum computing. Coherent control of the electrion-trion transitions in a single dot may be built upon to demonstrate two-photon control of the electron spin using the trion as an intermediate state \cite{WuPRL,EconomouPRL, YamNat}. In conjunction with spin control, pulsed optical read-out may also be used to evaluate the fidelities of quantum gate operations on a single spin qubit through density matrix tomography.

This work was supported in part by ARO, AFOSR, ONR, NSA/ARDA, DARPA and NSF.

\bibliography{APSRBpaper}

\begin{thebibliography}{29}
\expandafter\ifx\csname natexlab\endcsname\relax\def\natexlab#1{#1}\fi
\expandafter\ifx\csname bibnamefont\endcsname\relax
  \def\bibnamefont#1{#1}\fi
\expandafter\ifx\csname bibfnamefont\endcsname\relax
  \def\bibfnamefont#1{#1}\fi
\expandafter\ifx\csname citenamefont\endcsname\relax
  \def\citenamefont#1{#1}\fi
\expandafter\ifx\csname url\endcsname\relax
  \def\url#1{\texttt{#1}}\fi
\expandafter\ifx\csname urlprefix\endcsname\relax\def\urlprefix{URL }\fi
\providecommand{\bibinfo}[2]{#2}
\providecommand{\eprint}[2][]{\url{#2}}

\bibitem[{\citenamefont{Loss and DiVincenzo}(1998)}]{LossPRA}
\bibinfo{author}{\bibfnamefont{D.}~\bibnamefont{Loss}} \bibnamefont{and}
  \bibinfo{author}{\bibfnamefont{D.~P.} \bibnamefont{DiVincenzo}},
  \bibinfo{journal}{Phys. Rev. A} \textbf{\bibinfo{volume}{57}},
  \bibinfo{pages}{120} (\bibinfo{year}{1998}).

\bibitem[{\citenamefont{Imamoglu et~al.}(1999)}]{ImamogluPRL}
\bibinfo{author}{\bibfnamefont{A.}~\bibnamefont{Imamoglu}}
  \bibnamefont{et~al.}, \bibinfo{journal}{Phys. Rev. Lett.}
  \textbf{\bibinfo{volume}{83}}, \bibinfo{pages}{4204} (\bibinfo{year}{1999}).

\bibitem[{\citenamefont{Hanson et~al.}(2007)}]{HansonRMP}
\bibinfo{author}{\bibfnamefont{R.}~\bibnamefont{Hanson}} \bibnamefont{et~al.},
  \bibinfo{journal}{Rev. Mod. Phys.} \textbf{\bibinfo{volume}{79}},
  \bibinfo{pages}{1217} (\bibinfo{year}{2007}).

\bibitem[{\citenamefont{Saikin et~al.}(2008)}]{Saikinarx}
\bibinfo{author}{\bibfnamefont{S.~K.} \bibnamefont{Saikin}}
  \bibnamefont{et~al.}, \bibinfo{journal}{\texttt{arXiv:0802.1527v1}}
  (\bibinfo{year}{2008}).

\bibitem[{\citenamefont{Elzerman et~al.}(2004)}]{ElzermanNAT}
\bibinfo{author}{\bibfnamefont{J.~M.} \bibnamefont{Elzerman}}
  \bibnamefont{et~al.}, \bibinfo{journal}{Nature}
  \textbf{\bibinfo{volume}{430}}, \bibinfo{pages}{431} (\bibinfo{year}{2004}).

\bibitem[{\citenamefont{Kroutvar et~al.}(2004)}]{KroutvarNAT}
\bibinfo{author}{\bibfnamefont{M.}~\bibnamefont{Kroutvar}}
  \bibnamefont{et~al.}, \bibinfo{journal}{Nature}
  \textbf{\bibinfo{volume}{432}}, \bibinfo{pages}{81} (\bibinfo{year}{2004}).

\bibitem[{\citenamefont{Hanson et~al.}(2003)}]{HansonPRL}
\bibinfo{author}{\bibfnamefont{R.}~\bibnamefont{Hanson}} \bibnamefont{et~al.},
  \bibinfo{journal}{Phys. Rev. Lett.} \textbf{\bibinfo{volume}{91}},
  \bibinfo{pages}{196802} (\bibinfo{year}{2003}).

\bibitem[{\citenamefont{Petta et~al.}(2005)}]{JRPettaSCI}
\bibinfo{author}{\bibfnamefont{J.~R.} \bibnamefont{Petta}}
  \bibnamefont{et~al.}, \bibinfo{journal}{Science}
  \textbf{\bibinfo{volume}{309}}, \bibinfo{pages}{2180} (\bibinfo{year}{2005}).

\bibitem[{\citenamefont{Greilich et~al.}(2006)}]{BayerSCI}
\bibinfo{author}{\bibfnamefont{A.}~\bibnamefont{Greilich}}
  \bibnamefont{et~al.}, \bibinfo{journal}{Science}
  \textbf{\bibinfo{volume}{313}}, \bibinfo{pages}{341} (\bibinfo{year}{2006}).

\bibitem[{\citenamefont{Mikkelsen et~al.}(2007)}]{AwschNatPhys}
\bibinfo{author}{\bibfnamefont{M.~H.} \bibnamefont{Mikkelsen}}
  \bibnamefont{et~al.}, \bibinfo{journal}{Nature Physics}
  \textbf{\bibinfo{volume}{3}}, \bibinfo{pages}{770} (\bibinfo{year}{2007}).

\bibitem[{\citenamefont{DiVincenzo}(2000)}]{DiViFortsch}
\bibinfo{author}{\bibfnamefont{D.~P.} \bibnamefont{DiVincenzo}},
  \bibinfo{journal}{Fortschr. Phys.} \textbf{\bibinfo{volume}{48}},
  \bibinfo{pages}{9} (\bibinfo{year}{2000}).

\bibitem[{\citenamefont{Dutt et~al.}(2005)}]{GuruPRL}
\bibinfo{author}{\bibfnamefont{M.~V.~G.} \bibnamefont{Dutt}}
  \bibnamefont{et~al.}, \bibinfo{journal}{Phys. Rev. Lett.}
  \textbf{\bibinfo{volume}{94}}, \bibinfo{pages}{227403}
  (\bibinfo{year}{2005}).

\bibitem[{\citenamefont{Wu et~al.}(2007)}]{WuPRL}
\bibinfo{author}{\bibfnamefont{Y.}~\bibnamefont{Wu}} \bibnamefont{et~al.},
  \bibinfo{journal}{Phys. Rev. Lett.} \textbf{\bibinfo{volume}{99}},
  \bibinfo{pages}{097402} (\bibinfo{year}{2007}).

\bibitem[{\citenamefont{Berezovsky et~al.}(2008)}]{AwschSci}
\bibinfo{author}{\bibfnamefont{J.}~\bibnamefont{Berezovsky}}
  \bibnamefont{et~al.}, \bibinfo{journal}{Science}
  \textbf{\bibinfo{volume}{320}}, \bibinfo{pages}{349} (\bibinfo{year}{2008}).

\bibitem[{\citenamefont{Gammon et~al.}(1996)}]{GammonPRL}
\bibinfo{author}{\bibfnamefont{D.}~\bibnamefont{Gammon}} \bibnamefont{et~al.},
  \bibinfo{journal}{Phys. Rev. Lett.} \textbf{\bibinfo{volume}{76}},
  \bibinfo{pages}{3005} (\bibinfo{year}{1996}).

\bibitem[{\citenamefont{Leonard et~al.}(1994)}]{PetroffPRB}
\bibinfo{author}{\bibfnamefont{D.}~\bibnamefont{Leonard}} \bibnamefont{et~al.},
  \bibinfo{journal}{Phys. Rev. B} \textbf{\bibinfo{volume}{50}},
  \bibinfo{pages}{11687} (\bibinfo{year}{1994}).

\bibitem[{\citenamefont{Lee et~al.}(2001)}]{LeeAPL}
\bibinfo{author}{\bibfnamefont{H.}~\bibnamefont{Lee}} \bibnamefont{et~al.},
  \bibinfo{journal}{Appl. Phys. Lett.} \textbf{\bibinfo{volume}{78}},
  \bibinfo{pages}{105} (\bibinfo{year}{2001}).

\bibitem[{\citenamefont{Guest et~al.}(2002)}]{GuestPRB}
\bibinfo{author}{\bibfnamefont{J.~R.} \bibnamefont{Guest}}
  \bibnamefont{et~al.}, \bibinfo{journal}{Phys. Rev. B}
  \textbf{\bibinfo{volume}{65}}, \bibinfo{pages}{241310}
  (\bibinfo{year}{2002}).

\bibitem[{\citenamefont{Press et~al.}(2008)}]{YamNat}
\bibinfo{author}{\bibfnamefont{D.}~\bibnamefont{Press}} \bibnamefont{et~al.},
  \bibinfo{journal}{Nature} \textbf{\bibinfo{volume}{456}},
  \bibinfo{pages}{218} (\bibinfo{year}{2008}).

\bibitem[{\citenamefont{Zrenner et~al.}(2002)}]{ZrennerNat}
\bibinfo{author}{\bibfnamefont{A.}~\bibnamefont{Zrenner}} \bibnamefont{et~al.},
  \bibinfo{journal}{Nature} \textbf{\bibinfo{volume}{418}},
  \bibinfo{pages}{612} (\bibinfo{year}{2002}).

\bibitem[{\citenamefont{Ramsay et~al.}(2008)}]{AJPRL}
\bibinfo{author}{\bibfnamefont{A.~J.} \bibnamefont{Ramsay}}
  \bibnamefont{et~al.}, \bibinfo{journal}{Phys. Rev. Lett.}
  \textbf{\bibinfo{volume}{100}}, \bibinfo{pages}{197401}
  (\bibinfo{year}{2008}).

\bibitem[{\citenamefont{Kim et~al.}(2008)}]{DannyPRL}
\bibinfo{author}{\bibfnamefont{D.}~\bibnamefont{Kim}} \bibnamefont{et~al.},
  \bibinfo{journal}{Phys. Rev. Lett.} \textbf{\bibinfo{volume}{101}},
  \bibinfo{pages}{236804} (\bibinfo{year}{2008}).

\bibitem[{\citenamefont{Ware et~al.}(2005)}]{WarePRL}
\bibinfo{author}{\bibfnamefont{M.~E.} \bibnamefont{Ware}} \bibnamefont{et~al.},
  \bibinfo{journal}{Phys. Rev. Lett.} \textbf{\bibinfo{volume}{95}},
  \bibinfo{pages}{177403} (\bibinfo{year}{2005}).

\bibitem[{\citenamefont{Alen et~al.}(2003)}]{AlenAPL}
\bibinfo{author}{\bibfnamefont{B.}~\bibnamefont{Alen}} \bibnamefont{et~al.},
  \bibinfo{journal}{Appl. Phys. Lett.} \textbf{\bibinfo{volume}{83}},
  \bibinfo{pages}{2235} (\bibinfo{year}{2003}).

\bibitem[{\citenamefont{Xu et~al.}(2007)}]{XuPRL}
\bibinfo{author}{\bibfnamefont{X.}~\bibnamefont{Xu}} \bibnamefont{et~al.},
  \bibinfo{journal}{Phys. Rev. Lett.} \textbf{\bibinfo{volume}{99}},
  \bibinfo{pages}{097401} (\bibinfo{year}{2007}).

\bibitem[{\citenamefont{Patton et~al.}(2003)}]{WoggonPRB}
\bibinfo{author}{\bibfnamefont{B.}~\bibnamefont{Patton}} \bibnamefont{et~al.},
  \bibinfo{journal}{Phys. Rev. B} \textbf{\bibinfo{volume}{68}},
  \bibinfo{pages}{125316} (\bibinfo{year}{2003}).

\bibitem[{\citenamefont{Xu et~al.}(2008)}]{CPTXD}
\bibinfo{author}{\bibfnamefont{X.}~\bibnamefont{Xu}} \bibnamefont{et~al.},
  \bibinfo{journal}{Nature Physics} \textbf{\bibinfo{volume}{4}},
  \bibinfo{pages}{692} (\bibinfo{year}{2008}).

\bibitem[{\citenamefont{Burkard}(2008)}]{HoleCohNatMat}
\bibinfo{author}{\bibfnamefont{G.}~\bibnamefont{Burkard}},
  \bibinfo{journal}{Nature Materials} \textbf{\bibinfo{volume}{7}},
  \bibinfo{pages}{100} (\bibinfo{year}{2008}).

\bibitem[{\citenamefont{Economou and Reinecke}(2007)}]{EconomouPRL}
\bibinfo{author}{\bibfnamefont{S.~E.} \bibnamefont{Economou}} \bibnamefont{and}
  \bibinfo{author}{\bibfnamefont{T.~L.} \bibnamefont{Reinecke}},
  \bibinfo{journal}{Phys. Rev. Lett.} \textbf{\bibinfo{volume}{99}},
  \bibinfo{pages}{217401} (\bibinfo{year}{2007}).

\end{thebibliography}

\end{document}